# A cold-atom random laser

Q. Baudouin, N. Mercadier[†], V. Guarrera[‡], W. Guerin and R. Kaiser[*]

Institut Non Linéaire de Nice, CNRS, Université de Nice Sophia-Antipolis, 1361 route des Lucioles, 06560 Valbonne, France
[†]Present address: Saint-Gobain Recherche, 39 quai Lucien Lefranc, 93303 Aubervilliers, France
[‡]Present address: Research Center OPTIMAS, Technische Universität Kaiserslautern, 67663 Kaiserslautern, Germany
[*]e-mail: robin.kaiser@inln.cnrs.fr

**Conventional lasers make use of optical cavities to provide feedback to gain media. Conversely, mirrorless lasers can be built by using disordered structures to induce multiple scattering, which increases the effective path length in the gain medium and thus provides the necessary feedback[1]. These so-called random lasers[2-7] potentially offer a new and simple mean to address applications[6] such as lighting[8]. To date, they are all based on condensed-matter media. Interestingly, light or microwave amplification by stimulated emission occurs also naturally in stellar gases[9-11] and planetary atmospheres[12,13]. The possibility of additional scattering-induced feedback (that is, random lasing) has been discussed[11,14] and could explain unusual properties of some space masers[15]. Here, we report the experimental observation of random lasing in a controlled, cold atomic vapour, taking advantage of Raman gain. By tuning the gain frequency in the vicinity of a scattering resonance, we observe an enhancement of the light emission of the cloud due to random lasing. The unique possibility to both control the experimental parameters and to model the microscopic response of our system provides an ideal test bench for better understanding natural lasing sources, in particular the role of resonant scattering feedback in astrophysical lasers.**

A cloud of cold atoms constitutes a new medium to study random lasing, allowing a detailed microscopic understanding of gain and scattering. Multiple scattering of light in cold atoms has been extensively studied in the past[16,17]. Also, quasi-continuous lasing with cold atoms as gain medium, either placed inside optical cavities[18-21] or based on distributed feedback[22], has recently been demonstrated, illustrating the potential for a variety of gain mechanisms in a regime where optical coherence is limited by purely radiative decay channels. This is significantly different with respect to most random lasing devices, based on pulsed excitation of condensed matter systems, where the relaxation rates of the optical coherence are several orders of magnitude faster than the decay of the excited state population. Long phase coherence times however allow for efficient feedback by resonant scattering, as expected in astrophysical lasers[11].

To combine sufficient gain and scattering while using only one atomic species, we take advantage of the multilevel structure of rubidium atoms, shown in Fig. 1a (D$_2$ line of $^{85}$Rb, wavelength $\lambda$ = 780 nm). Two-photon Raman gain is obtained by a population inversion between the two hyperfine ground states |2> and |3> sustained by optical pumping. A Raman laser drives the |3> → |2'> transition with a large detuning $\Delta$, so that atoms can be transferred into the |2> state by stimulated emission. Scattering required for random lasing is provided by the |2> → |1'> line, which is a closed transition efficient for multiple scattering. Both Raman gain and scattering can occur at the same frequency (that is, for the same photons) by an appropriate choice of the Raman detuning $\Delta = -4.8\,\Gamma$, determined by the hyperfine splitting between the |1'> and |2'> states

($\Gamma/2\pi$ = 6 MHz is the linewidth of the transition). We thus tune the Raman laser to the vicinity of this condition and define the detuning $\delta = \Delta + 4.8\,\Gamma$ as the relevant parameter (Fig. 1a). This scheme takes advantage of the selection rules, which forbid electric dipole transitions between the states |3> and |1'>, so that the |1'> level does not affect the Raman gain.

For a given amount of gain and scattering, the threshold of random lasing is determined by a critical minimum size of the sample[1]. In our case, gain and scattering are provided by the same atoms and depend on the atomic density $n$. We have shown[23-25] that the critical parameter defining the random-laser threshold is the on-resonance optical thickness $b_0$, defined for a homogeneous cloud of radius $R$ as $b_0 = 2n\sigma_{34'}R$, with $\sigma_{34'}$ the on-resonance scattering cross-section for the |3> → |4'> transition (used to measure $b_0$, see Methods). Moreover, its critical value can be computed from the atomic polarizability only (see Supplementary Information).

Our sample consists of a cloud of cold $^{85}$Rb atoms collected in a magneto-optical trap. A controlled compression period provides a variable optical thickness $b_0$ with a constant number of trapped atoms (see Methods). We then switch off all lasers and magnetic field gradients during 1 ms before applying strong counterpropagating Raman beams (intensity $I_{Ra}$ = 4.25 mW/cm$^2$ per beam with crossed linear polarizations) tuned around $\delta \sim 0$. In addition, we use an optical-pumping laser tuned slightly below the |2> → |3'> transition to sustain a steady-state population inversion between the two hyperfine levels involved in our scheme. The relative intensity between the two external lasers allows us to adjust the relative populations, and thus to tune continuously from a sample with large gain and no scattering (with all atoms in the |3> state) to a situation without gain and with large scattering on the |2> → |1'> line (with all atoms in the |2> state). The data presented here have been obtained with an optical-pumping intensity $I_{OP}$ = 2.9 mW/cm$^2$, corresponding, at $\delta$ = 0, to 71% of atoms in the |3> state (evaluated for a single atom).

The signature of random lasing is obtained in our system by the detection of the total emitted light from the sample, which we collect with a solid angle of 10$^{-2}$ sr at an angle of 40° with respect to the Raman beam axes (Fig. 1b). Indeed, in contrast to other random-lasing devices, where the scattered pump light is orders of magnitude stronger than the random-laser emission itself, the random-laser line in our system has a strength comparable to the one of the other involved transitions. In addition, the external lasers are not scattered by the |2> → |1'> transition. Figure 2 shows the measured fluorescence as the Raman laser frequency is swept through the region of interest, for different values of the optical thickness $b_0$ of the atomic cloud (see Methods). We stress that for these measurements we vary $b_0$ while keeping the atom number constant. Variations in the fluorescence can thus only be related to collective features.

The first signature of such a collective behaviour can be seen in a regime of negligible scattering, far from the |2> → |1'> transition (regions 1 of Fig. 2): amplified spontaneous emission (ASE) induces an overall increase of the fluorescence as a function of $b_0$. Photons from the Raman beam can indeed undergo a spontaneous Raman transition. The subsequent scattered light is then amplified by Raman gain produced by the surrounding atoms while leaving the sample with a ballistic path. The efficiency of this process is directly related to the optical thickness (see Supplementary Information). The ASE signal decreases as the Raman laser is detuned further away from the |3> → |2'> transition (located at $\delta$ = +4.8 $\Gamma$) since both the spontaneous (source

contribution) and stimulated (gain contribution) Raman scattering rates decrease for larger detuning. Note that when tuning the Raman laser very close to the |3> → |2'> line, single-photon scattering dominates. As detailed in ref. 26, population redistribution is then responsible for the increase of fluorescence. This effect is negligible for the detunings considered here, and only gain can explain the observed features.

When the Raman laser is tuned close to $\delta = 0$ (region 2 of Fig. 2), the combination of gain and scattering gives rise to a random laser. It appears as an enhanced fluorescence bump that emerges as the optical thickness $b_0$ is increased. To better extract this signal, we fit the wings of the curves (regions 1) by adjustable slope and curvature and remove this ASE background. The remaining random-laser signal is a Gaussian peak, well-centred at $\delta = 0$ (Fig. 3a), which thus comes from the scattering due to the |2> → |1'> transition. Therefore, the observed peak is due to the combination of gain and scattering. Moreover, the peak amplitude shows a threshold behaviour, with a change of slope at $b_0 = 6\pm1$ (Fig. 3b). This threshold is the signature of the occurrence of random lasing in our sample when the Raman beams are tuned around $\delta \sim 0$ and when $b_0 > 6$. We stress that varying the optical thickness acts simultaneously on the amount of gain and feedback provided by the medium. This is unusual in laser physics, where the threshold is most-commonly defined as a critical pump power. In our case, increasing the optical-pumping intensity increases indeed the population inversion that provides gain, but simultaneously decreases the feedback, so that random lasing needs a fine tuning of the laser parameters.

Finally, we have exploited the possibility to perform *ab initio* theory by developing two simple models, one for ASE and the other for random lasing. Both are detailed in the Supplementary Information. Here, we describe briefly the random-lasing model. It consists in self-consistently coupling the atomic response, based on optical Bloch equations (OBE) with additional scattering on the |2> → |1'> line, to a diffusion equation for the light scattered on the |2> → |1'> resonance. The OBE allows us to compute the atomic polarizability and, including the additional scattering on the |2> → |1'> line, the mean-free path $\ell_{sc}$ and the gain length $\ell_g$, including saturation effects due to the random laser intensity $I_{RL}$ inside the sample. Like in conventional laser theory, we look for a steady-state solution where gain exactly compensates losses. In the diffusive regime and taking into account only the diffuse mode with the longest lifetime, this condition is equivalent to Letokhov's result on the random-lasing threshold[1,23],

$$R_{cr} = \pi \sqrt{\frac{\ell_{sc}\ell_g}{3}} \quad , \tag{1}$$

where $R_{cr}$ is the critical sample size. As a consequence, for a given $b_0$, we find the value of $I_{RL}$ such that Eq. (1) is fulfilled. Considering the simplicity of our model and the absence of adjustable parameter, the qualitative agreement between the experimental data and the computed values for the additional fluorescence is very satisfactory (Fig. 3b,c). The quantitative discrepancies suggest the need for more involved models. Many ingredients have indeed been neglected, like interference effects on light transport, light polarization, the Zeeman degeneracy of the involved atomic levels, the finite temperature, the inhomogeneous atomic density distribution, and cooperative effects[21,27]. The comparison between the experiment and new models including some of these effects will allow one to identify the most relevant ones and thus to better understand random-laser physics.

In summary, we have presented an experimental evidence of combined gain and scattering of light in a cloud of cold atoms, demonstrating random lasing in a dilute vapour. This type of experiments, based on well-controlled atomic systems, with the possibility of *ab initio* calculations, will allow the comparison to different models of random lasing[23,27-29] and help understanding complex situations encountered in astrophysical systems. Testing light propagation models with realistic atomic structures and designing novel detection schemes, based for instance on high-order photon correlations[30], are promising examples of such an approach.

## Methods

**Sample preparation.** In our experiment 6 counter-propagating trapping beams with a waist of 3.4 cm ($1/e^2$ radius of the intensity distribution) are used to load $^{85}$Rb atoms from a background atomic vapour into a magneto-optical trap (MOT). The trapping beams are detuned by $-3\Gamma$ from the $|3\rangle \rightarrow |4'\rangle$ hyperfine transition. To maintain the atomic populations in $|3\rangle$, we add 6 repumper beams tuned slightly below the $|2\rangle \rightarrow |3'\rangle$ transition. We can load between $10^8$ and $10^{11}$ atoms by changing the background vapour pressure and the duration of the trap loading (from 10 to 500 ms). Once the atoms are trapped in the MOT, we perform a temporal dark-MOT stage by increasing the detuning of the trapping beams to $-6\Gamma$ and by reducing the intensity of the repumper beams to few percents of their initial value. This leads to an increase of the spatial density and thus of the optical thickness $b_0$ of the cloud, without loss of atoms. By changing the duration of this compression stage, we are able to tune $b_0$ from 1.9 to 27, while keeping almost constant the total number of atoms, which, for the measurements presented here, is set to $7\times10^8 \pm 12\%$. The temperature is $T \sim 50$ µK. The optical thickness $b_0$ is measured by a transmission spectrum with a small and weak probe beam on the $|3\rangle \rightarrow |4'\rangle$ transition[24]. The shot-to-shot fluctuations of $b_0$ (horizontal error bars in Fig. 3b) are evaluated by repeating the measurement five times.

**Data acquisition.** After the sample preparation, we switch off magnetic field gradients and trapping lasers and we expose the sample to two counterpropagating Raman beams with waists of 2.4 cm and intensities of 4.25 mW/cm$^2$ each and 6 optical pumping beams with waists of 3.4 cm and intensities of 0.48 mW/cm$^2$ each at a detuning of $-2\Gamma$ from the $|2\rangle \rightarrow |3'\rangle$ transition. Note that the diameters of these lasers are large enough to insure that their effective intensities on the atom cloud are independent from the chosen optical thickness. The Raman laser is obtained from a distributed-Bragg-reflector laser diode and is frequency-tuned by a double-pass acousto-optic modulator before it is amplified by two stages of saturated slave lasers. This system allows us to scan the frequency in a range up to $16\Gamma$ with intensity variations of only 0.1%.

The measuring procedure consists in scanning in 2 ms the Raman beam detuning $\delta$ from $-3,2\Gamma$ to $4,8\Gamma$ while a high gain photodiode gathers the fluorescent emission of the cloud in a solid angle of $\sim 10^{-2}$ sr. We checked that the direction of the sweep does not change the detected fluorescence, and that the duration of the sweep is short enough to avoid significant variations of $b_0$ during the measurements (< 5%) and long enough to probe a quasi-steady state (the sweep rate is $4\Gamma$/ms). We averaged over 4000 subsequent measurements in order to increase the signal-to-noise ratio, thus performing also an averaging over the disorder configurations.

**Acknowledgments**
We acknowledge financial support from ANR (project ANR-06-BLAN-0096), CG06, PACA, DGA and the Research Executive Agency (program COSCALI, No. PIRSES-GA-2010-268717). We thank R. Carminati and S. Skipetrov for fruitful discussions, and A. Aspect and S. Tanzilli for useful comments on the manuscript.

**Author contributions**
Q.B., N.M. and V.G. made the experiment and analysed the data; Q.B., N.M. and R.K. developed the theory; Q.B., W.G. and R.K. wrote the paper; R.K. supervised the project. Q.B. and N.M. contributed equally to the study. All authors discussed the results and commented on the manuscript.

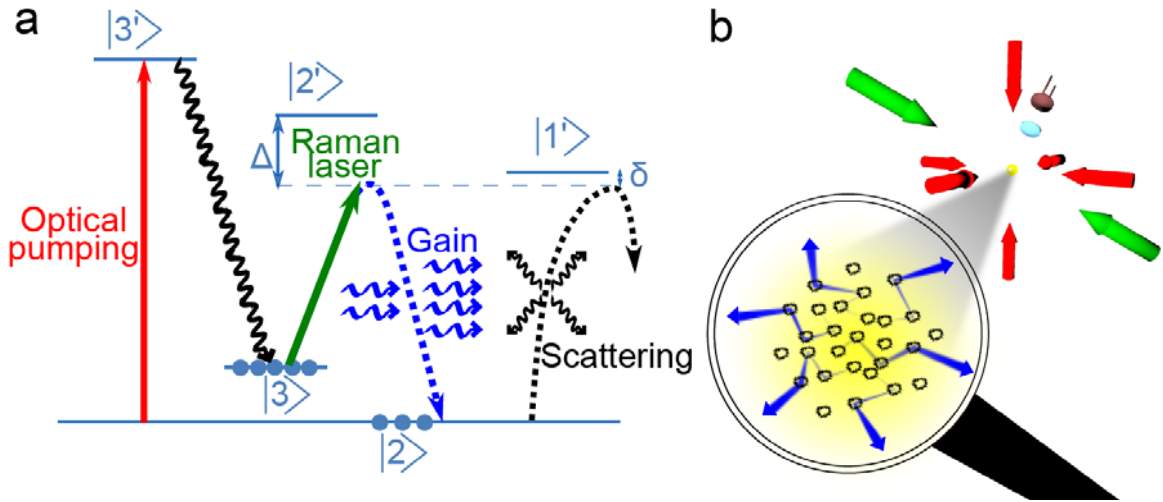

**Figure 1 | Working principle of the random laser.**
**a**, Atomic transitions of the $D_2$ line of $^{85}Rb$ used to create random lasing in cold atoms. The two hyperfine ground states are $|2\rangle = |F=2\rangle$ and $|3\rangle = |F=3\rangle$. Similarly, the involved hyperfine excited states are the states $|F''=1,2,3\rangle$ denoted $|1'\rangle$, $|2'\rangle$, and $|3'\rangle$. Optical pumping creates a population inversion between $|2\rangle$ and $|3\rangle$. This allows us to create Raman gain by applying a laser with a detuning $\Delta$ from the $|3\rangle \rightarrow |2'\rangle$ transition. The gain frequency has a detuning $\delta$ from the closed $|2\rangle \rightarrow |1'\rangle$ transition. Around $\delta \sim 0$, this transition provides efficient scattering. Random lasing can thus occur around this frequency.
**b**, Schematic representation of the experiment. The magneto-optical trap (yellow sphere) is exposed to two Raman-laser beams (green) and six optical-pumping beams (red). Its fluorescence is collected by a lens and detected by a photodiode. Zoom-in: light (in blue) is scattered by atoms in the $|2\rangle$ state (black) and amplified by atoms in the $|3\rangle$ state (yellow background).

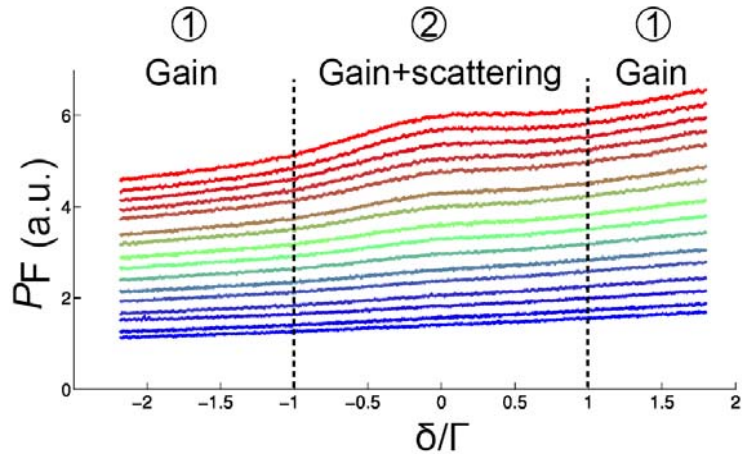

**Figure 2 | Fluorescence measurement.**
Total fluorescence $P_F$ as a function of the Raman laser detuning $\delta$ (in units of the linewidth $\Gamma$ of the optical transition) recorded for optical thickness varying from $b_0 = 1.9$ to $b_0 = 26$. The number of atoms is kept constant at $N = 7\times10^8 \pm 12\%$. Two collective features are visible. In the wings (regions 1), the overall increase of the fluorescence with $b_0$ is due to amplified stimulated emission. Around $\delta \sim 0$ (region 2), an additional peak appears for large optical thickness. This enhanced light emission is due to the combination of Raman gain and multiple scattering provided by the $|2\rangle \rightarrow |1'\rangle$ transition and is thus a signature of random lasing.

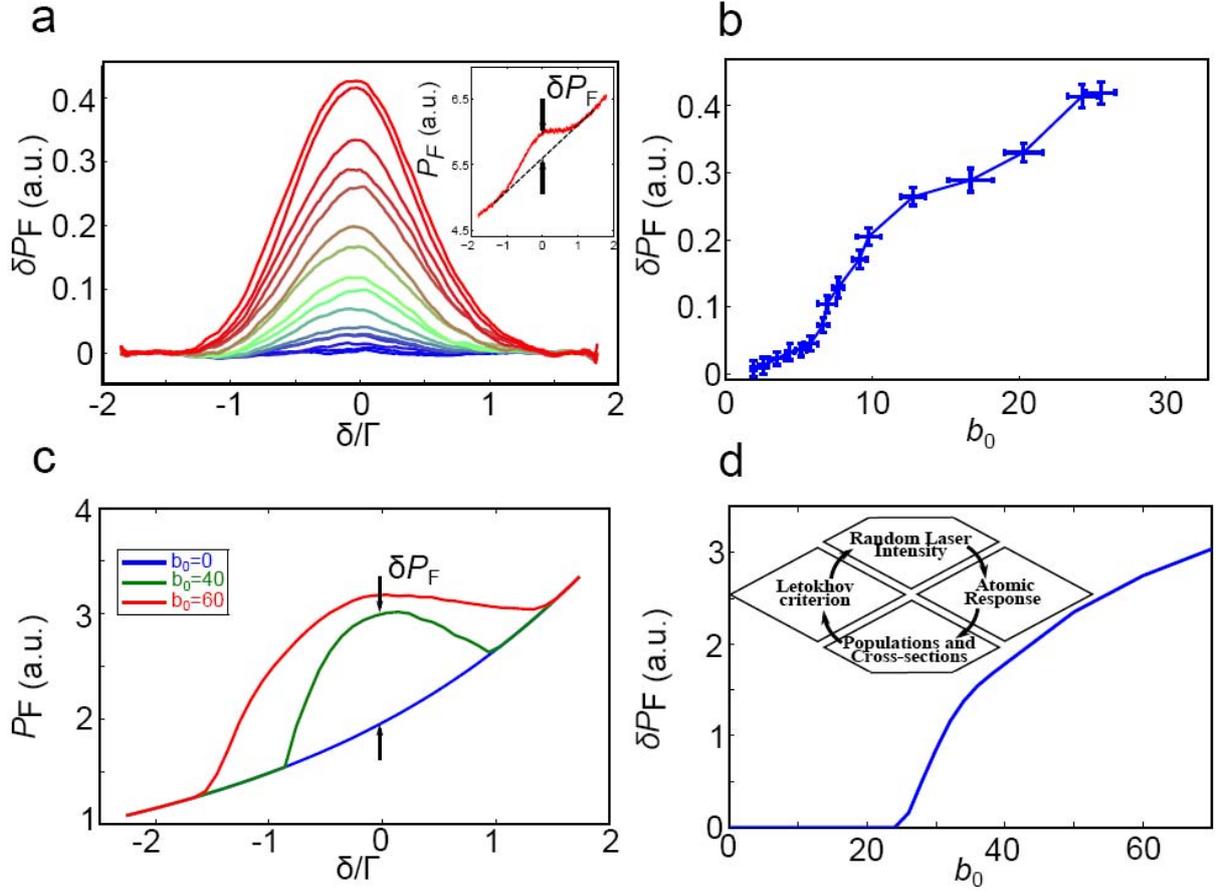

**Figure 3 | Random laser emission around δ = 0.**
**a**, Supplementary fluorescence $\delta P_F$ due to random lasing as a function of the Raman laser detuning δ, measured for optical thickness varying from $b_0 = 1.9$ to $b_0 = 26$. The source data are the same as in Fig. 2, but the wings (regions 1 in Fig. 2) have been subtracted. The inset illustrates the fitting procedure. For clarity we also smoothed the data, corresponding to a detuning resolution of 0.3 Γ.
**b**, Amplitude of the additional peak due to random lasing. A threshold optical thickness is clearly visible at $b_0 \sim 6$. Vertical error bars correspond to the rms noise on the data of Fig. 2 and horizontal error bars to shot-to-shot fluctuations of $b_0$.
**c**, Computed fluorescence $P_F$ as a function of the detuning δ with three different optical thicknesses ($b_0 = 0$ corresponds to the single-atom limit), using our self-consistent model for random lasing. It shows that an increase of the total fluorescence is expected around δ = 0 when random lasing occurs.
**d**, Solution at δ = 0 of the self-consistent model for random lasing. Only the supplementary fluorescence $\delta P_F$ is plotted, to allow for a comparison with the experimental data (panel **b**). Inset: principle of the model. The atomic response allows the computation of the threshold optical thickness following Letokhov's criterion[1,23]. Due to saturation effects, this threshold depends on the random-laser intensity. Therefore, for each $b_0$, we find the random-laser intensity such that the computed threshold equals $b_0$, corresponding to a steady state (see Supplementary Information).